\documentclass[runningheads]{llncs}
\usepackage{graphicx}
\usepackage{todonotes}
\usepackage{subcaption}

\begin{document}
\title{Preserving Tangible and Intangible Cultural Heritage: the Cases of Volterra and Atari}
\titlerunning{Preserving Tangible and Intangible Cultural Heritage Panel}
%
\author{
Maciej Grzeszczuk\inst{1,2}\orcidID{0000-0002-9840-3398} \and
Kinga Skorupska\inst{1}\orcidID{0000-0002-9005-0348} \and
Paweł Grabarczyk \inst{3} \orcidID{0000-0002-1268-7474} \and
Władysław Fuchs \inst{4,5}\orcidID{} \and
Paul F. Aubin  \inst{5}\orcidID{} \and
Mark E. Dietrick  \inst{5}\orcidID{0000-0002-9840-3398} \and
Barbara Karpowicz\inst{1}\orcidID{0000-0002-7478-7374} \and
Rafał Masłyk\inst{1}\orcidID{0000-0003-1180-2159} \and
Pavlo Zinevych \inst{i}\orcidID{0009-0008-9250-8712} \and
Wiktor Stawski \inst{1}\orcidID{0000-0001-8950-195X} \and
Stanisław Knapiński \inst{1}\orcidID{0009-0006-0524-2545} \and
Wiesław Kopeć \inst{1,6}\orcidID{0000-0001-9132-4171}
}
\authorrunning{Grzeszczuk et al.}
%
\institute{XR Center, Polish-Japanese Academy of Information Technology
 \url{https://xrc.pja.edu.pl}
\and
The Foundation for the History of Home Computers
\and
IT University of Copenhagen
\and
University of Detroit Mercy
\and
Volterra-Detroit Foundation
\and
Kobo Association}

\maketitle              
\begin{abstract}

At first glance, the ruins of the Roman Theatre in the Italian town of Volterra have little in common with cassette tapes containing Atari games. One is certainly considered an important historical landmark, while the consensus on the importance of the other is partial at best. Still, both are remnants of times vastly different from the present and are at risk of oblivion. Unearthed architectural structures are exposed to the elements just as the deteriorating signals stored on magnetic tapes. However, the rate of deterioration is much faster with the magnetic media, as their life expectancy is counted in decades, whereas the Roman Theater, which is already in ruin, measures its lifespan in centuries. Hence, both would benefit from some form of digital preservation and reconstruction. In this panel, we discuss how to sustainably preserve tangible and intangible cultural artifacts for future generations.

\keywords{Cultural Heritage \and Virtual Reality  \and Magnetic Media \and Heritage Preservation.}

\end{abstract}
\section{Rationale}

At first glance, the ruins of the Roman Theater in the Italian town of Volterra have little in common with cassette tapes containing Atari software. One is certainly considered an important historical landmark, while the consensus on the importance of the other is partial at best\footnote{On December 1, 2021, \textit{Demoscene - culture of demo creators} was included on the National List of Intangible Cultural Heritage - \url{https://kskpd.pl}}. Still, both are remnants of times vastly different from the present and are at risk of oblivion. Unearthed architectural structures are exposed to the elements just as the deteriorating signals stored on magnetic tapes. However, the rate of deterioration is much faster with the magnetic media, as their life expectancy is counted in decades, whereas the Roman Theater, which is already in ruin, measures its lifespan in centuries. Hence, both would benefit from some form of digital preservation and reconstruction, strengthening the bond between people and the past, and shaping perceptions, identity, environment, and residence of a society\cite{amali2022}. 

The mentioned tapes carrying bits of data are only a part of a greater group of intangibles like the practices, representations, expressions, knowledge or skills, which communities, groups, and sometimes individuals recognize as part of their cultural heritage\footnote{Convention for the Safeguarding of the Intangible Cultural Heritage: https://ich.unesco.org/en/convention}. If digitized in time, they can survive. Then they can be distributed in the form of identical copies to users around the world, who can undertake work to further analyze, describe, and preserve them\cite{amali2022}.

What if we can do the same with the ruins? If a group of people came with specialized scanners and made precise measurements and inventory of the archaeological site, they could continue its analysis also during unfavorable weather conditions, consult with specialists from the other side of the world, or finally distribute the work among a larger group of people, without the risk of trampling each other and the respected monument. Moreover, by periodically repeating such measurements, we can store snapshots of individual phases of the object's exploration and current state, enabling time travel to perform additional analysis, otherwise impossible due to the progress of work. In time, this representation could become the only remnant of the object, if the original is destroyed as a result of further decay, a cataclysm, or warfare\cite{nabiev2019}.

And this is where the 2000-year-old rocks start to travel together with the Atari - when digitized, the content needs further care, to achieve the preservation goal, however it is defined. Not all of it needs to be publicly accessible or presented right away, but it definitely needs to be properly organized, or else it will be lost as just another bucket of bits in the vast ocean of storage. At some point, the database with which it is stored (with the added structure, metadata, and products of the further work) becomes an artifact on its own. It starts to require continued curation, not only to maintain storage continuity\cite{storagestory2021}, but also consistency and relevance for the audience. Important questions start popping up, like how to ensure that these artifacts continue to be preserved? Are the existing preservation frameworks used in other projects sufficient in terms of sustainability and effectiveness? How to effectively broadcast the accumulated cultural heritage in an engaging, but informative, and aesthetically pleasing way to the public or interested scholars? And finally, are there regulatory guidelines to support this?\cite{kruglikova2020}

The development of technology helps in many ways. Storage, although far from perfect\cite{storagestory2021}, is becoming more affordable; new technologies, such as generative models, help fill the purely technical gaps, such as imperfections in the read or scanned data, or defects in the scanned material itself. But more importantly, they can fill the scene with emotions and impressions and create a whole new interactive experience\cite{kruglikova2020}. XR Headsets, which transfer audiences to virtual or extended reality, allow us to recreate historical scenes and locations\cite{VR3DHeritage2006,VR3DHeritage2020} that, brought to life with generative AI\cite{generativeAI}, could be teeming with life. Mobile device capabilities, improving every year, lead to "nomadic museography", where users themselves, with the use of their smartphones, can act as sources of data about cultural objects directly on their territory\cite{nabiev2019}, such as in the HeritageTogether project in Wales\cite{WalesHeritage3DCrowd_2015}. Even Wikipedia can be seen as a crowd effort to preserve knowledge and artifacts, as people can contribute images of objects in their possession, family pictures, and parts of personal narratives; their knowledge of familiar locations and historical events\cite{skorupska2020chatbot}. Transferring a part of the operational area to the virtual layer makes it easier to enrich not only online knowledge repositories, but also museum resources with individually created and curated collections, while ensuring their survival and continued availability in the event that there is no longer a person who has so far piloted the project independently\cite{behrendtz2021}.

    \begin{figure}[]
    \setkeys{Gin}{width=0.31\linewidth}
    \captionsetup[subfigure]{skip=0.5ex,
                             belowskip=1ex,
                             labelformat=simple}
    \renewcommand\thesubfigure{}
    \newcommand\sep{\hspace{0.025\linewidth}}

    \subfloat[Figure 1: Live Bulletin Board System access made available to participants of the retrocomputing event in 2022 as a demonstration of what networking looked like in the 1980s.]{\includegraphics{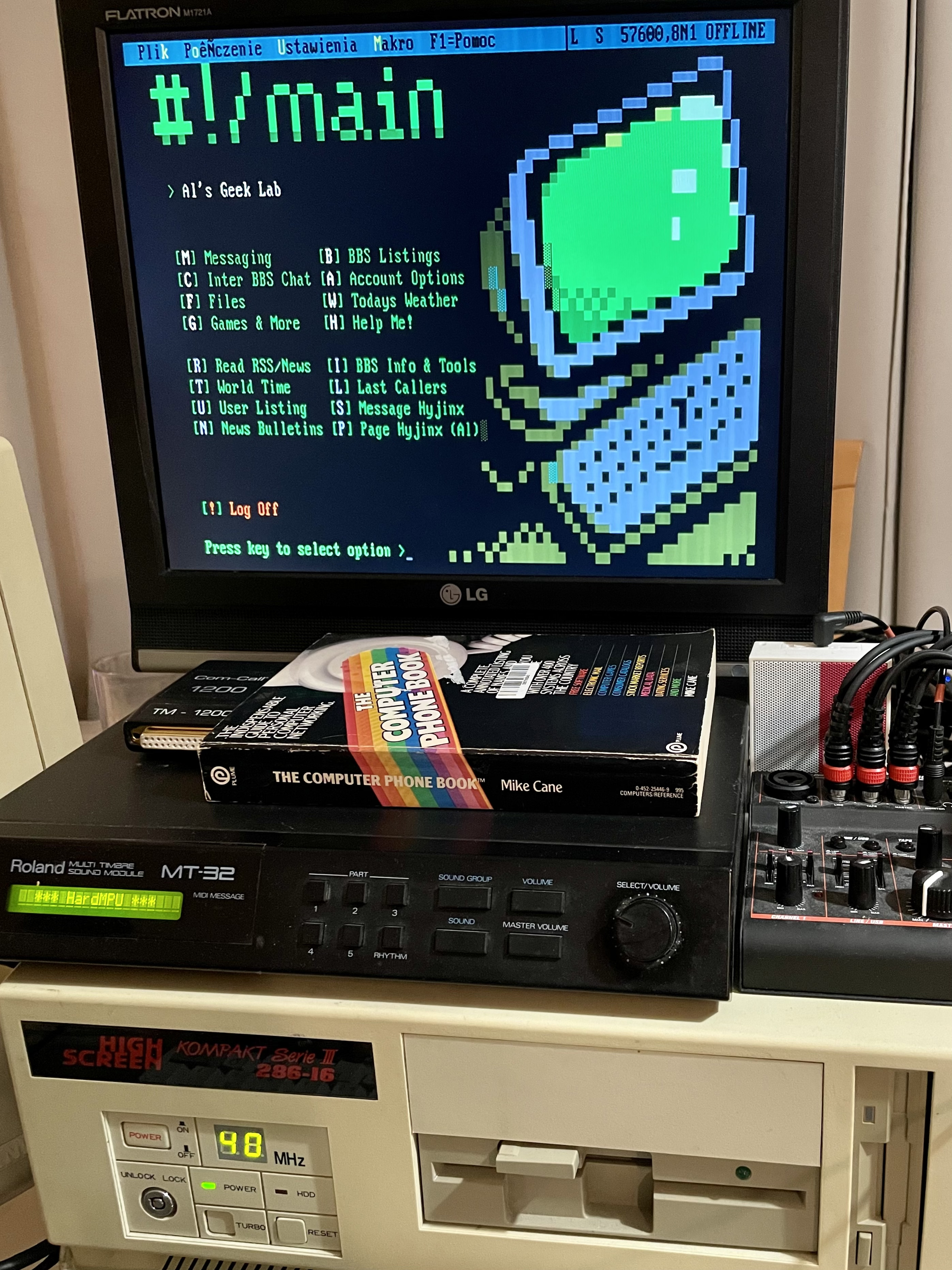}}
    \sep
    \subfloat[Figure 2: Turbo Soft's Videocartridge, unique Chilean interface from 1980s. Allowed using VHS video tapes as a high-capacity software storage for 8-bit Atari. Demonstration on Lost Party 2022.]{\includegraphics{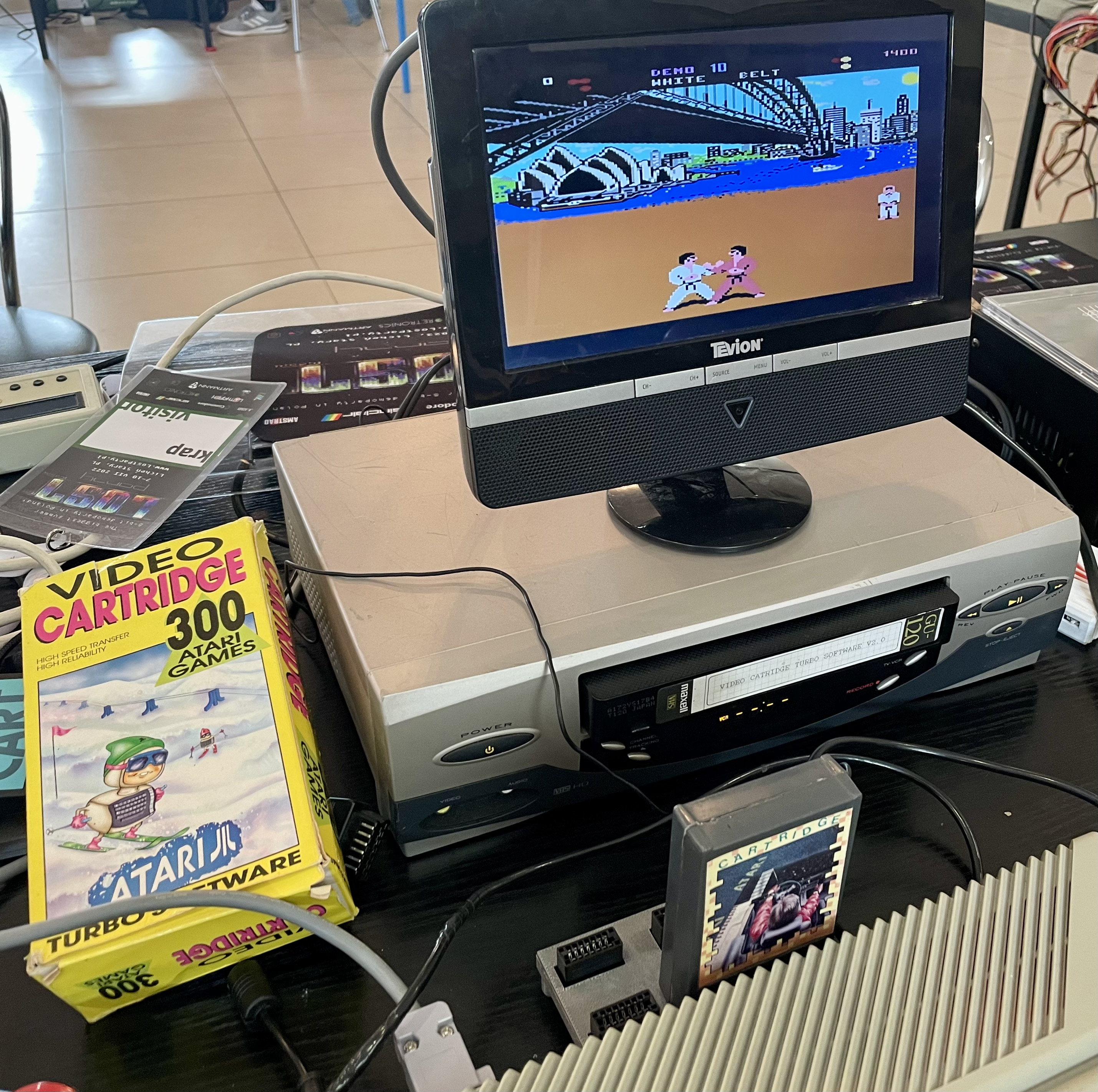}}
    \sep
    \subfloat[Figure 3: Magnetic tapes degrade not only over time, but also as a result of mechanical damage that may occur during normal use.]{\includegraphics{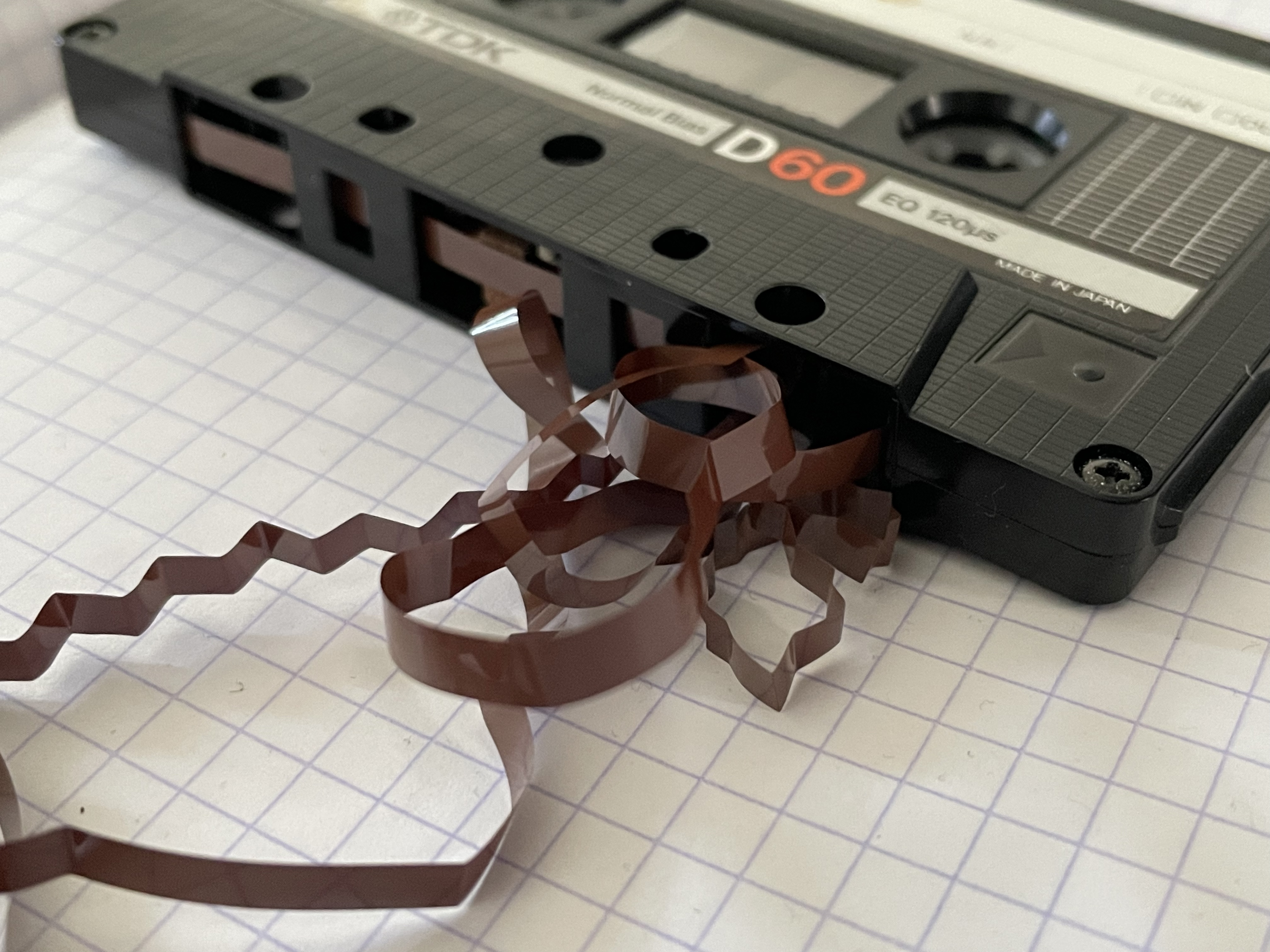}}
    
    \medskip
    \subfloat[Figure 4: Only a handful of units of the prototype Atari 1090 expansion module have survived to this day. Here it is as presented at VCF East 2019.]{\includegraphics{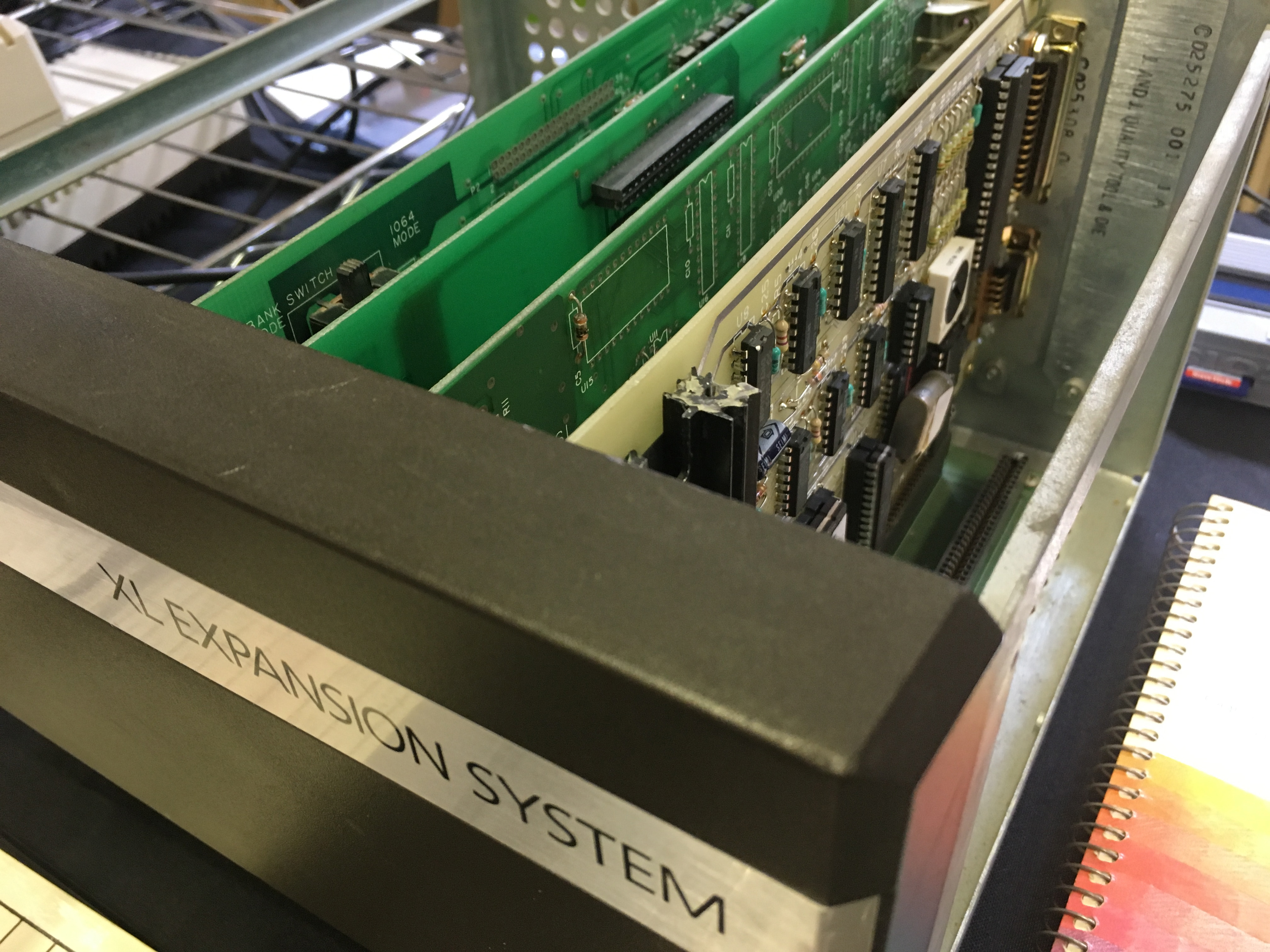}}
    \sep
    \subfloat[Figure 5: The photo shows the main screen of "Top Secret BBS" with a stylized ANSI image of its operator, Marcin Borkowski, running in DOSBOX - a DOS emulator.]{\includegraphics{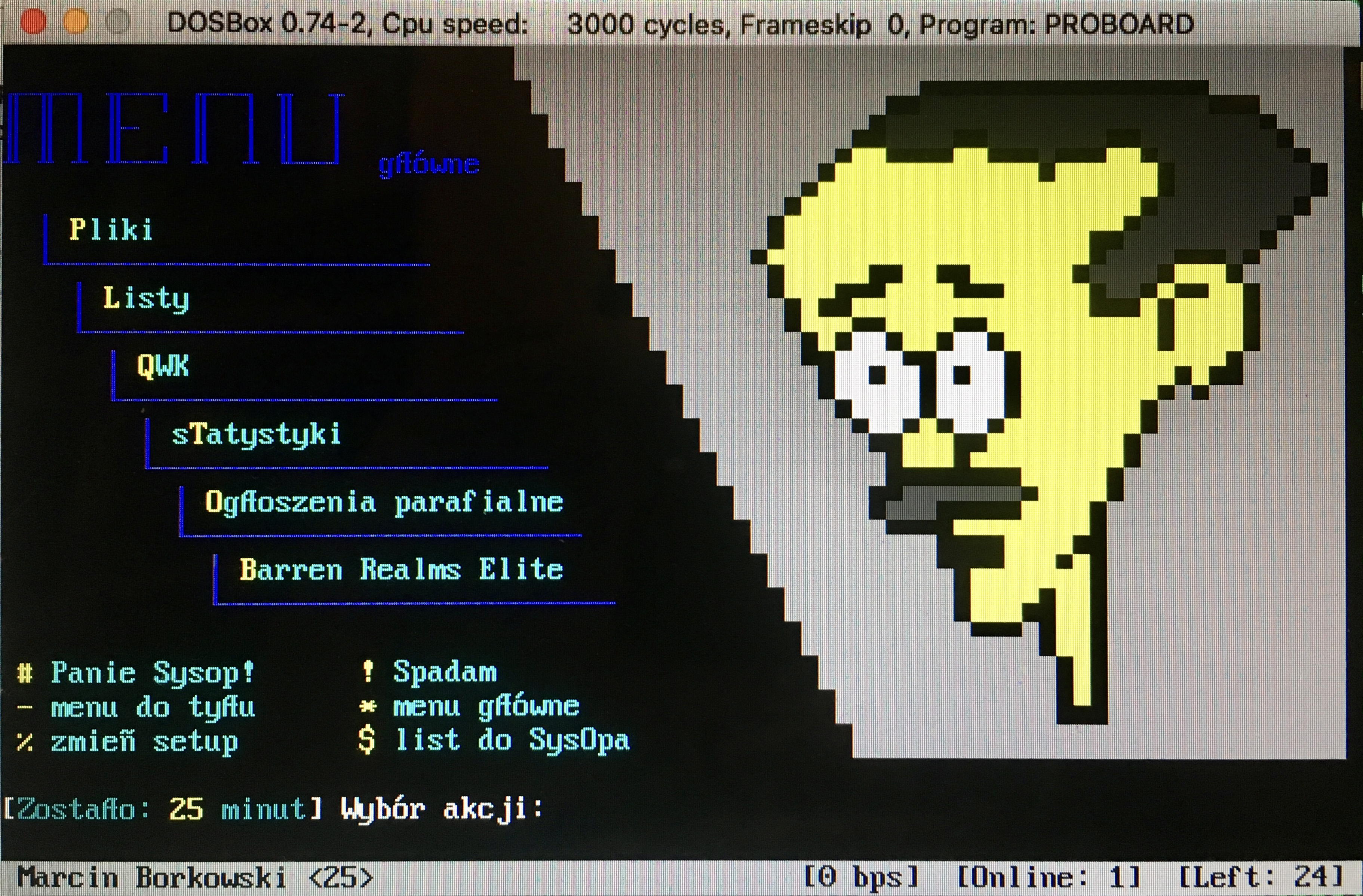}}
    \sep
    \subfloat[Figure 6: A real-time rendering of a fragment of an ancient Roman theater in Volterra, using Unreal Engine 5 and dynamic Lumen lighting.]{\includegraphics{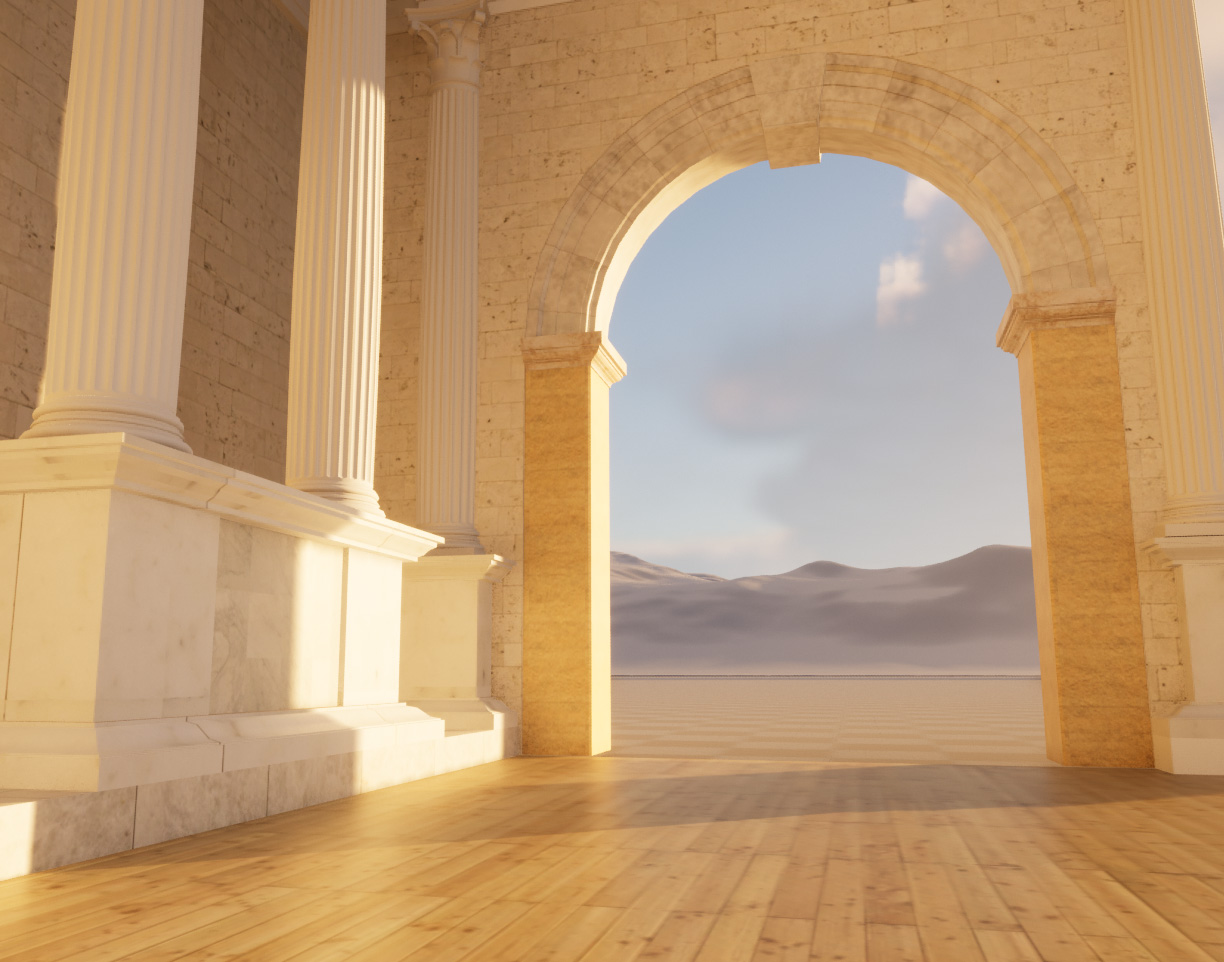}}

    \medskip
    \subfloat[Figure 7: A real-time rendering of the ancient Roman theater scene in Volterra, using Unreal Engine 5 based on the artistic vision of Władysław Fuchs.]{\includegraphics{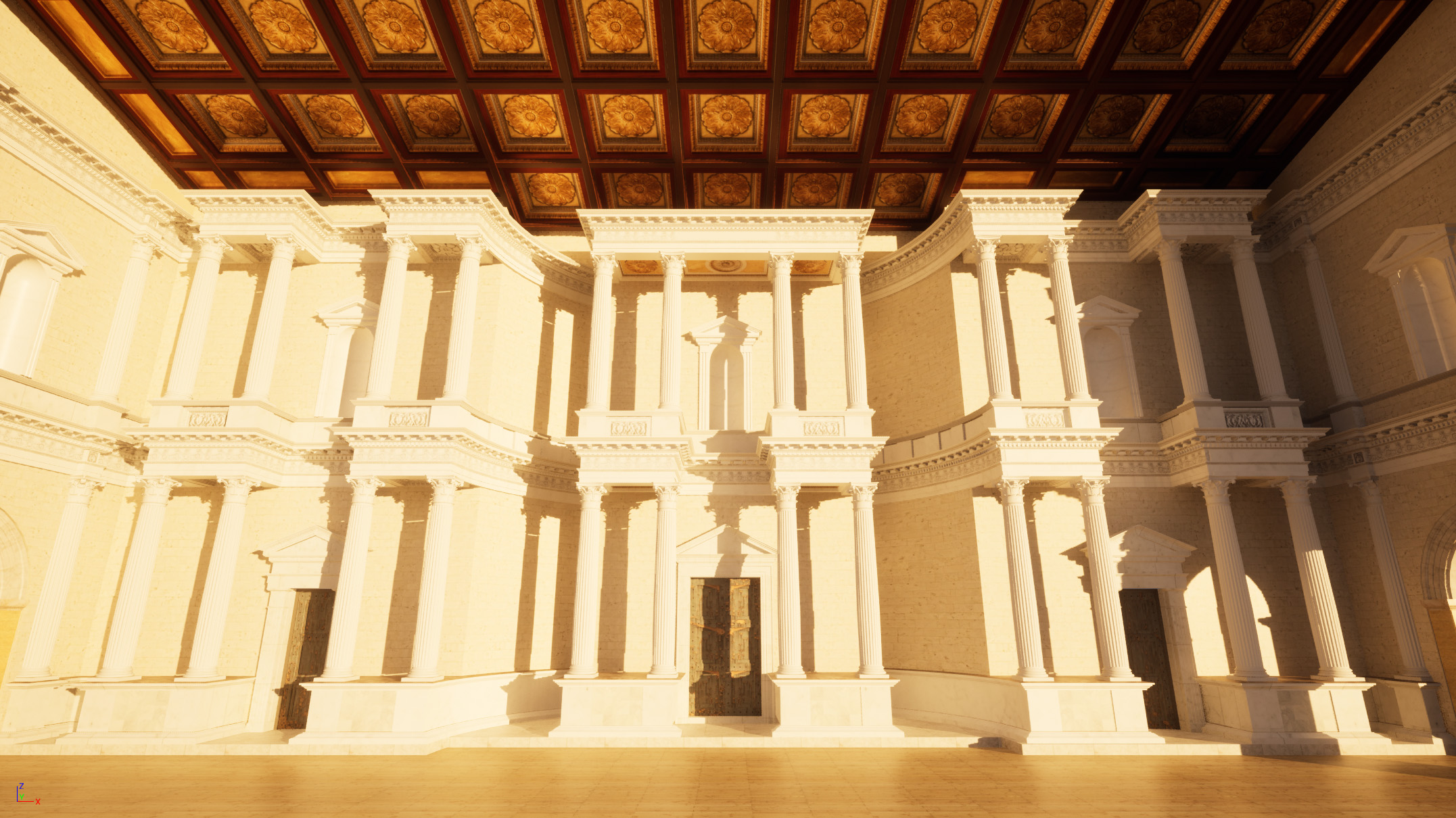}}
    \sep
    \subfloat[Figure 8: Reconstruction of the column based on materials collected at the ancient theater in Volterra, made with the Unreal Engine 5 engine.]{\includegraphics{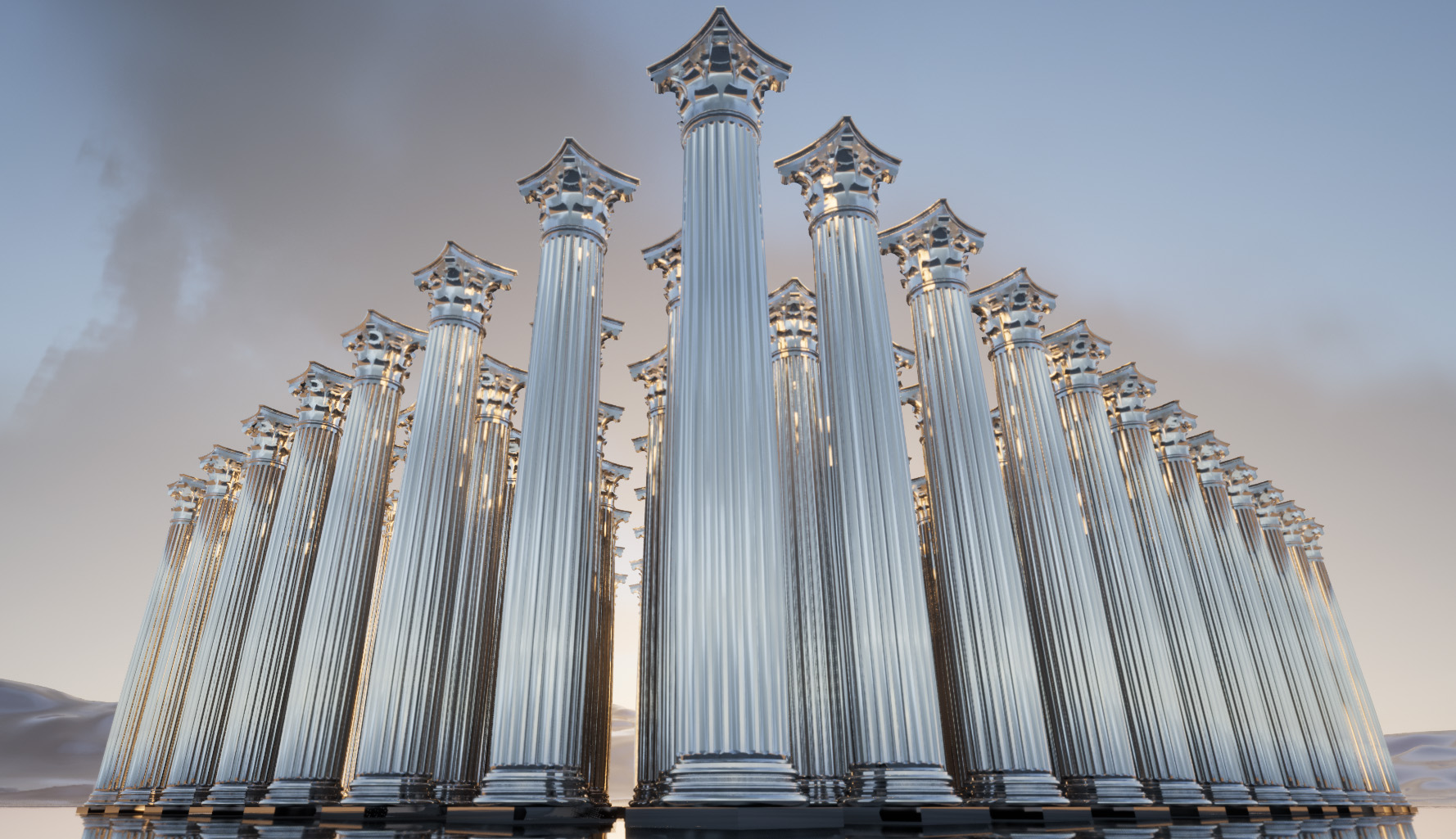}}
    \sep
    \subfloat[Figure 9: Reconstruction of the ancient theater in Volterra in Unreal Engine 5 showcasing capabilities of the Nanite tool, optimizing the 3D model in real time.]{\includegraphics{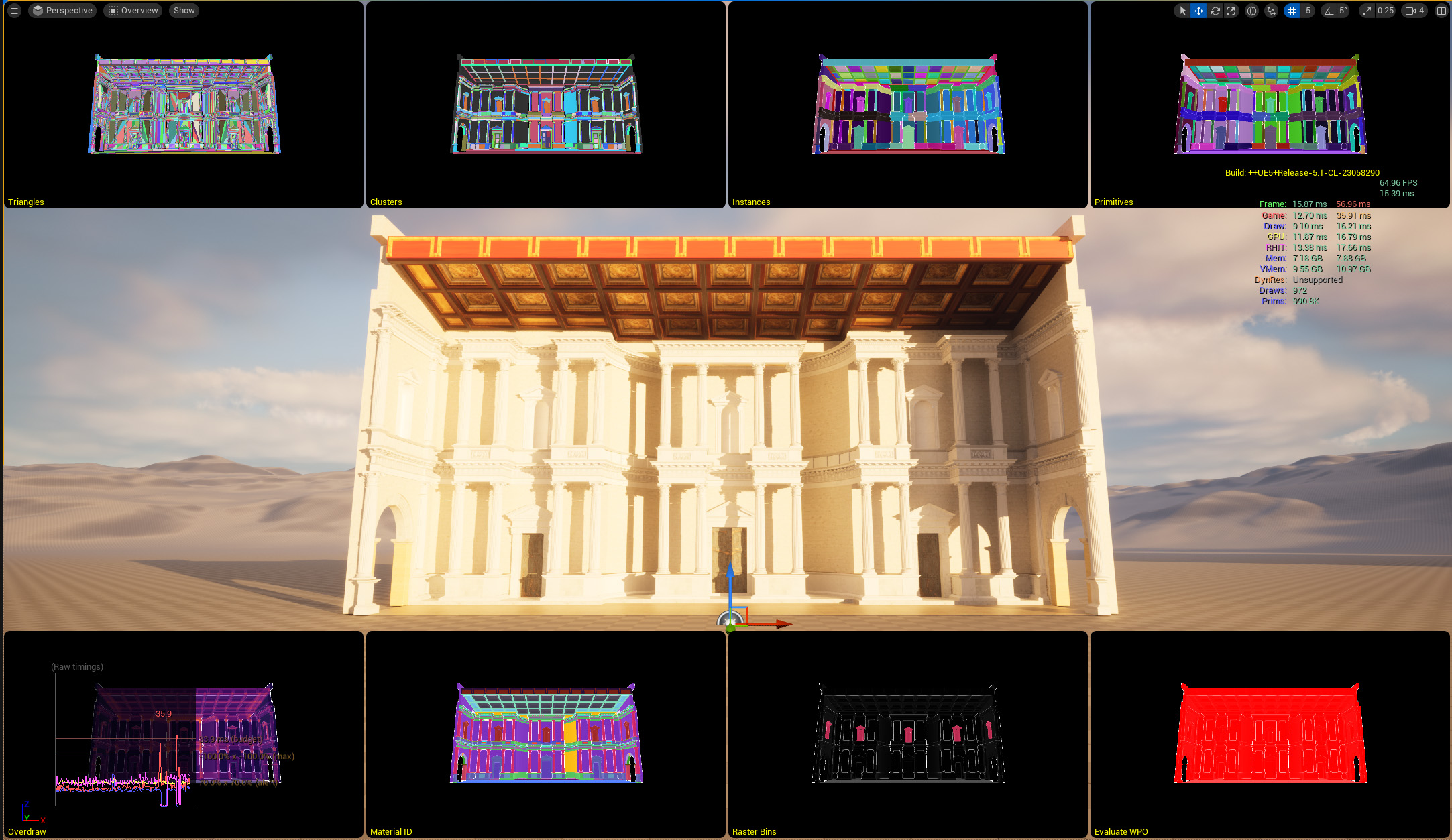}}

    \end{figure}

\section{Theme}

During the panel, we will discuss what we mean by the preservation of tangible and intangible cultural heritage, what goals we want to achieve through our projects, and how we want to achieve them. We will consider whether working with historical artifacts that are so different in form, age, and function provides a chance to use some of the proven solutions, or to avoid certain already known pitfalls? 

\subsection{Focus}

Taking advantage of the fact that the panelists represent a wide range of fields and have extensive experience within their areas of interest, ranging from the artifacts of the early digital age (see Figures 1-5) to 3D scanning and modeling physical historical sites (see Figures 6-9), we will start with a retrospective, identifying significant areas of project activities to date and sharing thoughts on what contributed positively, what negatively and how such challenges could be counteracted in future projects of a similar type. We will attempt to identify common difficulties and bottlenecks and find inspiration from the know-how of the different disciplines for our own practice.

\subsection{Discussion Points}
\paragraph{Aspect 1: Transforming data into engaging narratives}

How to find out, what do audiences actually want to find out and experience? To what extent should scholars shape the experience with education in mind? 
What is the path from the idea, through data collection, to creating a memorable and engaging experience for the project's audience? 
Is the path well delineated from the beginning, and if not, how does it wind, and how does it affect the implementation of the project?

\paragraph{Aspect 2: Harnessing the power of communities}

How and when is it best to engage communities when undertaking these types of endeavors? What support can we expect, and how can we prepare to make the most of their potential? What could a sustainable community-supported preservation drive look like? How to build communities of practice that continue to not only aid the digitization efforts, but also support long-term preservation?

\paragraph{Aspect 3: Legislation: a challenge or an ally for long-term preservation?}

How can we sustainably achieve long-term preservation, and can we standardize storage formats and media to facilitate it? Are we legally empowered to carry out digitization and have the resulting digitized data, and in what areas can we then use it? What about privacy? Taking into account the further development of culture in its digital area, should we lobby for any legislative changes, and if so, what should they be and at what level (local, national, international)?

\section{Organizers}

\subsection{Panelists}

\paragraph{Władysław Fuchs}
Received his Master of Architecture degree in 1987 from Warsaw Institute of Technology. In 1994, he completed and defended his Ph.D. dissertation at the same institution. He has been awarded the Warsaw Institute of Technology President's Award. From 1987 until 1991 he worked full-time in the Free-hand Drawing Department of the School of Architecture in Warsaw. Since 1991 he has been working continuously at the School of Architecture, University of Detroit Mercy. 

\paragraph{Paweł Grabarczyk} Associate professor and head of the Games Research Group at the IT University of Copenhagen. By training, he is a philosopher of language and works on topics that connect philosophy and games studies, such as game ontology and the cultural significance of digital play. He is also interested in game history, game preservation and demoscene (currently working on a historical book on Atari 8-bit platform).

\paragraph{Mark Dietrick}
Throughout his career, Mark has focused on technology and research in support of innovation in practice and has excelled at balancing dedication to architecture with a nearly innate understanding of technology. As one of the first pioneers in the school of Computer Automated Design, Mark has been involved with the latest technology as it involves architecture, engineering, and construction. Mark received his Architectural degree from the University of Detroit Mercy in 1984. He is an Adjunct Faculty member at the University of Pittsburgh and the University of Detroit Mercy School of Architecture. He is a founding Board Member for the Volterra-Detroit Foundation and leads the Foundation’s Reality Computing research initiatives.

\paragraph{Wiesław Kopeć}
Computer scientist, research and innovation team leader, associate professor at Computer Science Faculty of Polish-Japanese Academy of Information Technology (PJAIT). Head of XR Center PJAIT and XR Department. He is also a seasoned project manager with a long-term collaboration track with many universities and academic centers, including University of Warsaw, SWPS University, National Information Processing Institute, and institutes of Polish Academy of Sciences. He co-founded the transdisciplinary HASE research group (Human Aspects in Science and Engineering) and distributed LivingLab Kobo.

\paragraph{Kinga Skorupska}
Assistant professor at the Polish-Japanese Academy of Information Technology doing research at the intersection of ICT and Social Sciences. Kinga's interests include online communities and collaboration, user experience, motivation, and games. She does research on ICT applications for social good, ranging from inclusion of diverse populations in the main technological discourse, through education to well-being.

\paragraph{Maciej Grzeszczuk}
Data scientist with a diverse professional background, including telecommunications, business processes, aviation, and air traffic control. Member of the Department of XR and Immersive Systems at the Polish-Japanese Academy of Information Technology, researching issues related to the human factor in extreme conditions and archiving old magnetic media in order to preserve cultural heritage. Privately, a fan of retrocomputers, sometimes a pilot, and more often a sailor and traveler.

\subsection{Researchers network}
We would like to thank the many people and institutions gathered by the Kobo Living Lab and the HASE Research Group to allow for this collaboration and research. First, we thank all the members of HASE research group (Human Aspects in Science and Engineering) and Living Lab Kobo for their support. In particular, the members of XR Center Polish-Japanese Academy of Information Technology (PJAIT) as well as its students' club, supporting Volterra preservation efforts, in particular Wiktor Stawski and Stanisław Knapiński as well as XR Center Staff involved in this process, that is Barbara Karpowicz, Rafał Masłyk and Pavlo Zinevych, and Emotion-Cognition Lab SWPS University (EC Lab), Kobo Association, Living Lab Kobo community, VR and Psychophisiology Lab of the Institute of Psychology Polish Academy of Sciences, The Foundation for the History of Home Computers, and National Information Processing Institute (NIPI). We would also like to thank the members of the Volterra-Detroit Foundation, in particular Władysław Fuchs, Paul F. Aubin and Mark E. Dietrick for their support of the topic of cultural heritage preservation in VR and beyond.

\section{Discussion and Conclusions}
Immersive environments and digital representations of physical objects and media became not only a means of presentation, but also of preservation and increased access. These \textbf{representations}, such as a Virtual Roman Theater in Volterra, \textbf{allow researchers to further analyze resulting digital objects, experience them in depth, and create new knowledge}. Such digitized activities can be independent of physical location, can be shared and distributed between the interested members, potentially worldwide; extending engagement far beyond groups of specialists or local communities. 

This comes with its own set of problems. Digital artifacts created as part of such activities \textbf{should themselves be subject to curation, documentation, and preservation} for future generations. The lack of standards and structure among involved communities, which often consist of enthusiasts with minimal formal or organizational support, lead to situations where objects of preservationist initiatives, sometimes containing results of impressive studies and massive collaborations, turn out to be insufficiently preserved, for example on a single person's server, and disappear forever. This touches on the issue of sustainability and commitment. Even highly effective preservation and \textbf{crowdsourcing initiatives require animation to function}. A project run by one person or stimulated by one person is at risk of disappearance when the leader is gone or runs out of motivation. Hence, distributing both data and responsibility among larger groups is a form of increasing the chances of preservation. 

Realizing the importance of protecting architectural monuments, which were built to last for centuries, took generations. In the case of intangible heritage - media containing unique data such as unique local releases of pirated software from the 1980s, environments to run software for early VR headsets and VR software from the 1990s or even source code of old Flash games - we don't have that much time. \textbf{Those who may have these artifacts in their possession may not even realize their importance}. 

Some areas may also be difficult to address without appropriate legislative support, since intellectual property rights protect even the products that are no longer within the scope of interest of their owners, if these entities exist at all. Scanning buildings or their facades that are at risk of perishing due to ongoing warfare, even in order to preserve their 3D representation, may also conflict with the rights to the architectural design. There are no regulations, as in the case of printed magazines or book publications, that would require the archiving of copies in national libraries, or a law requiring equipment manufacturers to offer spare parts, to allow repairs, or provide access to service manuals for older hardware. Therefore, \textbf{urgent legislative activities supporting digital cultural heritage would be necessary}, either that of purely digital origin, as well as digital representations of the physical objects. For this purpose, it seems important to increase public awareness and popularization activities. 

An important factor is having the ultimate goal in mind, which rarely is just a cloud of scanned points, or bits of data in the archive. To engage people, \textbf{we need to create a complete experience, use the object to tell a meaningful story}, interact with the needs of the audience, and build these experiences to be relevant and fascinating.

\bibliographystyle{splncs04}
\bibliography{bibliography}

\end{document}